\documentclass{article}

\usepackage{arxiv}

\usepackage[utf8]{inputenc} 
\usepackage[T1]{fontenc}    
\usepackage{hyperref}       
\usepackage{url}            
\usepackage{booktabs}       
\usepackage{amsfonts}       
\usepackage{nicefrac}       
\usepackage{microtype}      
\usepackage{lipsum}		
\usepackage{graphicx}
\usepackage{doi}

\usepackage{adjustbox}

\title{Beyond Tradition: Evaluating Agile feasibility in DO-178C for Aerospace Software Development}

\author{ \href{https://orcid.org/0000-0002-1894-3993}{\includegraphics[scale=0.06]{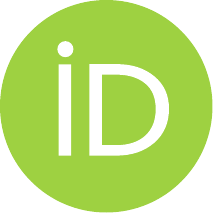}\hspace{1mm}J. Eduardo Ferreira Ribeiro} \\
	Department of Informatics Engineering \\
	Faculty of Engineering, University of Porto \\
	Porto, Portugal \\
	\texttt{jose.eduardo.ribeiro@fe.up.pt} \\
	\And
	\href{https://orcid.org/0000-0002-4800-5201}{\includegraphics[scale=0.06]{orcid.pdf}\hspace{1mm}João Gabriel Silva} \\
	CISUC, Department of Informatics Engineering \\
	University of Coimbra \\
	Coimbra, Portugal \\
	\texttt{jgabriel@dei.uc.pt } \\
	\And
	\href{https://orcid.org/0000-0002-4046-4729}{\includegraphics[scale=0.06]{orcid.pdf}\hspace{1mm}Ademar Aguiar} \\
	Department of Informatics Engineering \\
	Faculty of Engineering, University of Porto \\
	Porto, Portugal \\
	\texttt{ademar.aguiar@fe.up.pt} \\
}

\renewcommand{\shorttitle}{\textit{Beyond Tradition: Evaluating Agile feasibility in DO-178C for Aerospace SDP}}

\hypersetup{
pdftitle={Beyond Tradition: Evaluating Agile feasibility in DO-178C for Aerospace Software Development},
pdfsubject={cs.SE},
pdfauthor={J. Eduardo Ferreira Ribeiro, João Gabriel Silva, Ademar Aguiar},
pdfkeywords={Safety-critical systems, Aerospace, DO-178C, Software development, Agile},
}

\begin{document}
\maketitle

\begin{abstract}
Domain-specific standards and guidelines play a crucial role in regulating safety-critical systems, with one notable example being the \emph{DO-178C} document for the aerospace industry. This document provides guidelines for organisations seeking to ensure the safety and certification of their software systems. This paper analyses the \emph{DO-178C} document within the context of software development for safety-critical aerospace systems focusing on Agile software development, aiming to assess its feasibility. Unlike restricting specific development methods, \emph{DO-178C} offers indispensable support that upholds confidence in safety, aligning seamlessly with the objectives of aerospace industries. Our analysis reveals that there are no limitations or restrictions within the \emph{DO-178C} that inhibit the adoption of Agile and provides guidelines and objectives for achieving suitable evidence, allowing for various working methods, including Agile methods, contrary to the overall opinion in the industry that the traditional waterfall method is mandatory. Additionally, we emphasise that the guidelines explanation is explicitly tailored to software professionals using Agile methods, giving it a much more specific focus than publications that only provide a generic overview of the standard.
\end{abstract}


\section{Introduction}
\label{introduction}
The intrinsic link between safety and risk, as highlighted by Bowen et al. \cite{Bowen1993},  serves as the foundation for the concept of safety-critical systems. These systems, as defined by Cullyer \cite{Cullyer1993}, encompass those whose functional failure could result in dire consequences such as loss of life, injury, environmental damage, or property loss. Within the aerospace domain, where the role of software is pivotal, ensuring a high degree of reliability becomes paramount. Embedded software, a vital component of safety-critical systems, collaborates with hardware to manage overall equipment safety. System and software development processes are deeply intertwined in this intricate landscape, accentuating the need for robust safety assurance. Examples of safety-critical systems abound, encompassing aircraft flight control, medical devices, rail control, satellites, and more \cite{Knight2002}. The horizon of safety-critical systems broadens further as innovations like Tesla's self-driving technology take centre stage \cite{Cullyer1993, Myklebust2018}. As such, the software development methodologies underpinning these systems require meticulous consideration. Traditionally, the waterfall method has favoured the safety-critical systems engineering domain.

Additionally, the Verification \& Validation Model (V-Model), an evolved version of the waterfall method, has gained traction due to its incorporation of Verification and Validation (V\&V) phases \cite{Mathur2010}. Proposed by Paul Rook in the late 1980s, the V-Model continues to influence safety-critical systems engineering \cite{Mathur2010, Russo2016}. However, standards and documents like \emph{DO-178C} \cite{DO-178C} do not prescribe a specific development model and provide guidelines and objectives for achieving suitable evidence, allowing for various working methods, including Agile methods \cite{DO-178C, Rierson2013, VanceHildermanandTonyBaghai2007}. This paper establishes the groundwork by providing essential background information on safety-critical systems and introduces a pivotal document in aerospace software development, known as \emph{DO-178C} \cite{DO-178C}. In the safety-critical domain, including the aerospace domain, the systems must adhere to stringent regulatory requirements and guidelines to ensure reliability and safety. The \emph{DO-178C} document serves as a cornerstone, offering comprehensive guidance for organisations involved in software development and certification for aerospace applications. Within the scope of this paper, we embark on an analysis of \emph{DO-178C} within the context of software development for safety-critical aerospace systems, with a specific emphasis on Agile software development. Our primary goal is to evaluate its feasibility. It is worth noting that the influence of \emph{DO-178C} extends beyond its prescribed methods as an indispensable supporter in investing confidence in safety measures across the aerospace domain.

Furthermore, this paper conducts a comprehensive synthesis of \emph{DO-178C} tailored for software professionals who employ Agile methods within the context of Agile software development. We assess its feasibility in this context. This approach provides a significantly sharper focus compared to existing publications, often providing only a generic overview of the document.

Section \ref{aerospacestandardsanddocuments} offers an overview of the interrelationships among various aerospace standards and documents within the aerospace domain. This section places particular emphasis on the \emph{DO-178C} document. Moving on to Section \ref{do178history}, we delve into a series of subsections that commence with insights into the evolution of \emph{DO-178}, from its initial publication to its most recent version, \emph{DO-178C}, issued in 2011. Subsection \ref{failureconditioncategorylevel} introduces and explores the distinct Failure Condition Category Levels outlined in \emph{DO-178C}. This subsection succinctly summarizes the requisite objectives and verification objectives for each level, as stipulated by the \emph{DO-178C} document. Subsequently, Subsection \ref{do178candsupplements} highlights the central \emph{DO-178C} document and explains its connections to relevant supplementary documents. Within Subsection \ref{do178cstructure}, the \emph{DO-178C} document structure is introduced and explored. Moving on to Subsection \ref{DO-178CFivePlans}, we investigate the five essential plans mandated by \emph{DO-178C} for all levels. In Subsection \ref{SoftwareDevelopmentStandardsandPlansOutputs}, we delve into the correlations linking plans, processes, standards, and outputs that collectively facilitate adherence to \emph{DO-178C}'s document and for project-specific needs. Subsection \ref{SoftwareDevelopmentProcess} subsequently navigates the software development process, outlining the essential steps from initial planning to attaining certification-ready status. The outcome of this exploration is a visual representation depicting the interplay between these processes, the designated plans, and the objectives of safety certification. Within Subsection \ref{TraceabilityIndependenceandChangeControl}, we underscore the pivotal role of traceability, independence, and change control in successfully pursuing certification. Finally, in the same subsection, we offer a condensed insight into the objectives and activities of each process, as delineated across different levels of criticality. In Subsection \ref{do178cactivities}, we provide a comprehensive overview of the processes associated with the tables listed in Annex A of the \emph{DO-178C} document pointing to the location where these tables are situated within the \emph{DO-178C} document. We conclude with Section \ref{conclusions}, reflecting on the \emph{DO-178C} document's influence on aviation software safety, and we address the question of whether \emph{DO-178C} actually imposes a particular development model or only provides guidelines and objectives for acquiring sufficient evidence, thus enabling a variety of working methods, including Agile approaches. Additionally, we identify future research areas, explicitly evaluating established methods and exploring novel approaches to optimise efficiency, safety, and compliance within aerospace software practices.

\section{Aerospace Standards and Documents}
\label{aerospacestandardsanddocuments}
\begin{figure}[htbp]
\centerline{\includegraphics[width=0.6\textwidth]{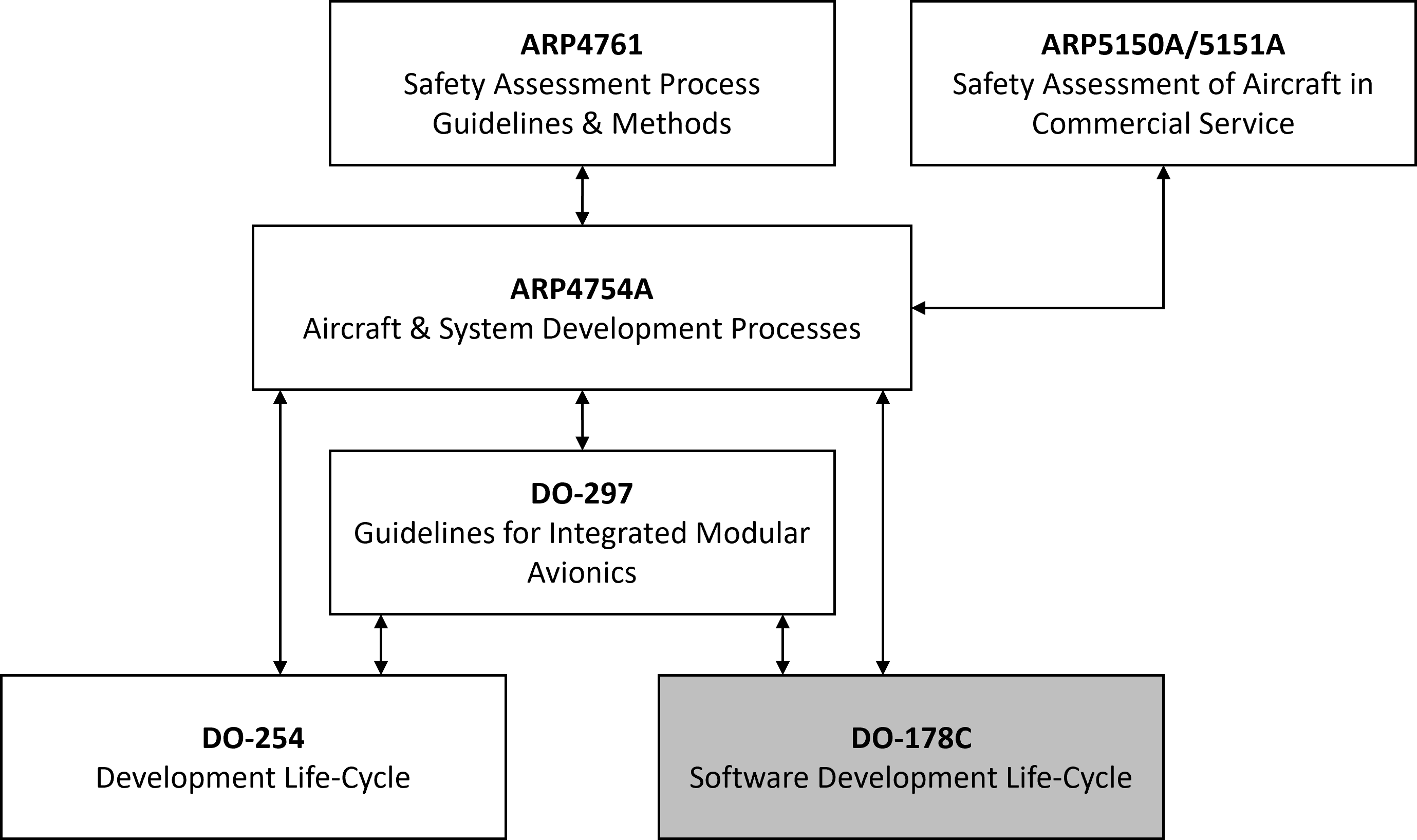}}
\caption{Safety-critical standards for aerospace relationships, adapted from \cite{ARP4754A}.}
\label{fig1}
\end{figure}

 Figure \ref{fig1} outlines the relationships of primary standards and documents in the aerospace domain. It highlights the relationships among various aerospace standards and documents, with \emph{DO-178C} and \emph{DO-254} \cite{DO-254} being the most widely recognised. Figure \ref{fig1} also provides a comprehensive structure that includes additional standards and documents at the system level, addressing various aspects of system development and certification.

By examining Figure \ref{fig1}, we can observe how these primary standards are linked. It highlights their pivotal role in guiding safety assessment, electronic hardware and software lifecycle processes, and the overall system development process. Each of these development documents provides essential guidelines to ensure the safety and reliability of aerospace systems.

In particular, while Figure \ref{fig1} shows the broader landscape of standards and documents in the aerospace domain, this specific analysis focuses mainly on the \emph{DO-178C} document. This document, also known as "Software Considerations in Airborne Systems and Equipment Certification," is of utmost importance in studying avionics system development and certification. It sets forth the guidelines and requirements that developers must adhere to when designing and implementing software for airborne systems.

Turning our attention back to Figure \ref{fig1}, we can see how it presents a concise overview of the interrelationships between key standards governing avionics systems. The diagram explicitly highlights the connections among the following standards:

\begin{itemize}
    \item \textbf{\emph{ARP4761}} focusses on the assessment of system safety, providing guidelines to identify and mitigate potential hazards in avionic systems \cite{ARP4761}.
    \item \textbf{\emph{ARP4754A}} deals with system development, addressing the activities and processes involved in the design, development, and validation of aircraft and systems \cite{ARP4754A}.
    \item \textbf{\emph{DO-297}}, known as the "Integrated Modular Avionics (IMA) Development Guidance and Certification Considerations," offers guidance for developing integrated modular avionics systems \cite{DO-297}.
    \item \textbf{\emph{DO-254}}, or "Design Assurance Guidance for Airborne Electronic Hardware," specifies the objectives and activities necessary for designing and verifying electronic hardware used in airborne systems \cite{DO-254}.
    \item \textbf{\emph{DO-178C}}, as mentioned above, sets the standards for software development assurance, ensuring that software for airborne systems is developed and tested to meet strict safety and reliability requirements \cite{DO-178C}.
    \item \textbf{\emph{ARP5150A/5151A}} focusses on electromagnetic environmental effects, guiding the management and mitigation of the impact of electromagnetic interference on avionic systems \cite{ARP5150A, ARP5151A}.
\end{itemize}

By adhering to these standards and documents, aerospace developers and certification authorities can work together to create robust and reliable avionics systems that meet the industry's strict requirements.

\section{DO-178 and its Evolution}
\label{do178history}
In the past, software held a more straightforward role, primarily enhancing the functions of mechanical and analogue electrical systems. However, an important realization surfaced early on: this approach could not ensure the reliability and safety required for intricate safety systems. This epiphany catalysed the release of the inaugural \emph{DO-178} certification document in 1980, as cited by Singh \cite{Singh2011}.

During this era, \emph{DO-178} established a form of software certification grounded in "best practices," although its content remained abstract. The regulations were gradually refined through trial and error. \emph{DO-178} introduced the concept of software verification requirements, with its specifics contingent upon the software's safety criticality. This document also categorized software applications into critical, essential, and nonessential sectors. Additionally, it intertwined the software certification process with relevant Federal Aviation Regulations (FARs), such as Type Certification Approval, Technical Standard Order (TSO) Authorization, and Supplemental Type Certification \cite{Singh2011}.

The maiden version of \emph{DO-178} laid the foundation for software certification. However, practical application swiftly necessitated revisions. As new needs emerged, more frequent updates ensued to enhance the guidelines. Table \ref{evolution_do_178} visually illustrates the progression from \emph{DO-178} to its latest incarnation, \emph{DO-178C}.

\begin{table}[htbp]
\caption{\emph{DO-178} evolution, adapted from \cite{DO-178C, VanceHildermanandTonyBaghai2007, McHale2001}.}
\begin{center}
\begin{adjustbox}{max width=\columnwidth} 
\begin{tabular}{|c|c|c|c|}
\hline
\textbf{Document} & \textbf{Basis} & \textbf{Year} & \textbf{Themes} \\
\hline
{\emph{DO-178}} & {\emph{498 \& 2167A}} & {1980} & {Artefacts, documents, traceability, management and testing.} \\
\hline
{\emph{DO-178A}} & {\emph{DO-178}} & {1985} & {Process, testing, components, four critical levels, reviewers, waterfall method.} \\
\hline
{\emph{DO-178B}} & {\emph{DO-178A}} & {1992} & {Integration, development methods, data (not documents), tools COTS.} \\
\hline
{\emph{DO-178C}} & {\emph{DO-178B}} & {2011} & {Reducing subjectivity, address modeling and reverse engineering.} \\
\hline
\end{tabular}
\end{adjustbox}
\label{evolution_do_178}
\end{center}
\end{table}

The 1982 publication of \emph{DO-178} established a "prescriptive set of design assurance processes for airborne software that focuses on documentation and testing" \cite{Singh2011}. This standard underwent numerous refinements, each aimed at clarifying and advancing concepts. Iterations introduced novel ideas when necessary, exemplified by the supplements in \emph{DO-178C}. \emph{DO-178}'s 1985 update, version A, introduced the concept of varied activities based on software component criticality (Subsection \ref{failureconditioncategorylevel}). The subsequent version, B, introduced the concept of activities and associated objectives, leading to a comprehensive overhaul and fostering development method flexibility in 1992. This iteration also delineated essential attributes a design assurance process must possess \cite{Singh2011}.

The most recent version, \emph{DO-178C}, emerged in 2011, retaining much of its predecessor's content while striving to clarify application nuances and eliminate inconsistencies. \emph{DO-178C} also introduced the notion of supplements, supplemental documents addressing cutting-edge technology's application without altering the core standard (Subsection \ref{do178candsupplements}). The \emph{DO-178} family of standards has long been viewed as a cornerstone of aviation safety. Literature affirms that its utilization has not been linked to significant aviation accidents \cite{Singh2011}.

In summary, the journey from the early days of software's role in mechanical enhancement to the intricate certification processes of \emph{DO-178C} showcases the evolution of software's integration into safety-critical systems within the aerospace domain.

\subsection{Failure Condition Category Level}
\label{failureconditioncategorylevel}
When developing an aerospace system, it is essential to identify the specific category of associated failure conditions. This initial comprehension holds enormous importance in ensuring strict adherence to the mandated standards, documents and levels the certifying authority sets during the validation process; otherwise, our systems will not be certified.

\begin{figure}[ht]
    \centering
    \includegraphics[width=.45\textwidth]{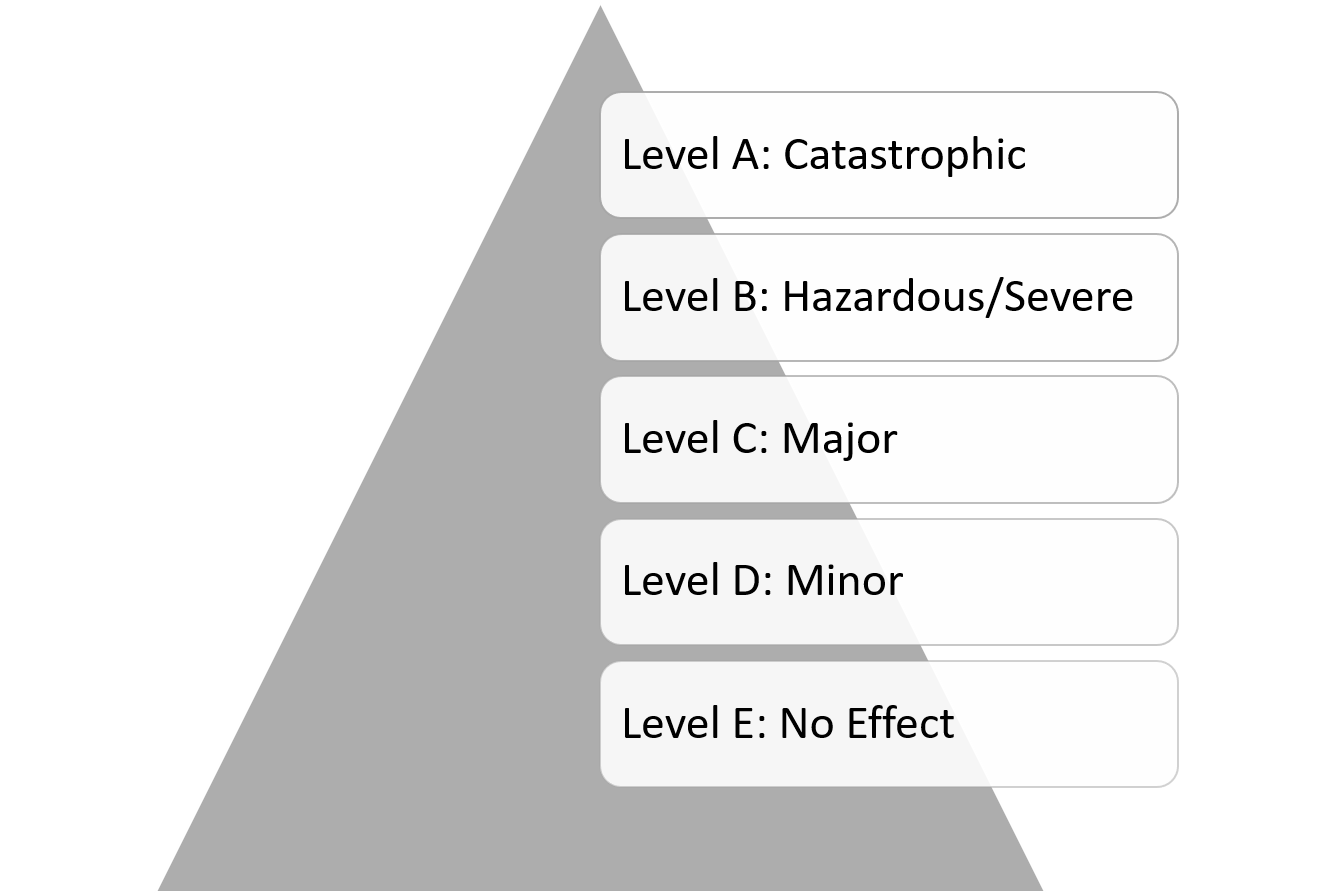}
    \caption{DAL pyramid, adapted from \cite{VanceHildermanandTonyBaghai2007}.}
    \label{fig2}
\end{figure}

The \emph{DO-178C} document demonstrates confidence in a software component in proportion to its designated failure condition category level, often called its criticality. To achieve this, the \emph{DO-178C} categorizes the criticality of components into five distinct Design Assurance Levels (DAL), as shown in Figure \ref{fig2}. This classification system acts as the foundation for determining the level assigned to a software component, considering its contribution to the potential failure conditions of the overarching system \cite{DO-178C}.

Below we have detailed each DAL level caused by software's anomalous behaviour determining or contributing to a failure of a system function \cite{DO-178C, VanceHildermanandTonyBaghai2007}.

\begin{itemize}
    \item \textbf{Level A: Catastrophic} is a failure condition resulting in multiple fatalities and loss of the aeroplane.
    \item \textbf{Level B: Hazardous/Severe} is a failure condition that reduces the aeroplane or flight crew's capability to cope with adverse operating conditions. These effects can manifest as a large reduction of functional capabilities and safety margins, as fatal or serious injury to a small number of passenger besides the crew and can create physical distress or excessive workload of the flight crew that can no longer accurately or completely do their tasks.
    \item \textbf{Level C: Major} is a failure condition that reduces the aeroplane or flight crew's capability to cope with adverse operating conditions. These failures can create a significant reduction of functional capabilities, safety margins, discomfort or distress of the flight crew and passengers possibly including injuries or can significant increase in the crews workload impairing their efficiency.
    \item \textbf{Level D: Minor} failure condition would not cause significant effect on the aeroplane safety, all crew members being in the possession of  knowledge needed to manage such incidents. These conditions may refer to slight reduction of safety margins and functional capabilities, slight increase in the crew workload as well as some physical discomfort to passengers and cabin crew.
    \item \textbf{Level E: has no effect} his a criticality level that does not impact the safety margins and would not affect on the operational capabilities of the aircraft. While this error does not affect the functioning of the aircraft, justifications about this error (issue) need to be provided to the Federal Aviation Administration (FAA).
\end{itemize}

The \emph{DO-178C} categorizes the DAL levels based on the required objectives, the impact of anomalies in the software component, and the failure rate of that component. These objectives encompass process requirements outlined in the document, showcasing adherence to \emph{DO-178C} when interacting with certification authorities. Table \ref{tab4} summarises these levels' overview, from the utmost criticality (level A) to the least (level E). As depicted in Figure \ref{fig2} and Table \ref{tab4}, level A is the highest criticality level according to the document. It requires a comprehensive set of objectives and verification objectives to ensure the safety and reliability of the software. Notably, level E, positioned as the least critical, is the only level excused from requiring verification activities. We might still need to follow certain development practices if our software falls under level E. However, the scrutiny and verification level is generally lower than the other levels.

\begin{table}[htbp]
\caption{Number of Objectives and Verification Objectives required by \emph{DO-178C} for every DAL level \cite{DO-178C}.}
\begin{center}
\begin{adjustbox}{max width=\columnwidth} 
\begin{tabular}{|c|c|c|c|}
\hline
\textbf{DAL} & \textbf{No. of Objectives } & \textbf{No. of Verification Objectives} \\
\hline
{A} & {71} & {43} \\
\hline
{B} & {69} & {41} \\
\hline
{C} & {62} & {34} \\
\hline
{D} & {26} & {9} \\
\hline
{E} & {0} & {0} \\
\hline
\end{tabular}
\end{adjustbox}
\label{tab4}
\end{center}
\end{table}

Presenting all objectives and verification objectives for each DAL summarised in Table \ref{tab4} is not feasible due to the extensive and complex information involved. For a thorough compilation of these objectives and verification objectives, referring directly to the \emph{DO-178C} document is recommended. This document provides comprehensive guidance on the requirements for each level, spanning from level A to the final level E. In Subsection \ref{do178cactivities}, we provide a comprehensive overview of the processes associated with the tables listed in Annex A of the \emph{DO-178C} document. This section precisely points to the location where these tables are situated within the \emph{DO-178C} document. These tables thoroughly explain the objectives, verification objectives, activities, and outcomes linked to distinct software levels.

The \emph{DO-178C} document expands beyond failure conditions and outlines additional software-related considerations within the system lifecycle process \cite{DO-178C}. These considerations encompass various aspects:

\begin{itemize}
    \item \textbf{Parameter Data Items} - Consists of executable object code and/or data that can compromise one or more configuration items or data that can influence the software behaviour without modifying the executable.
    \item \textbf{User-Modifiable Software} - Consists of software or part that may be changed by the user within modifications constraints without certification authority review.
    \item \textbf{Commercial-Off-The-Shelf Software} - Software already included in the airborne systems.
    \item \textbf{Option-Selectable Software} - Consists of functions from some airborne systems and equipment that may be selected by other software instead of hardware connector pins.
    \item \textbf{Field-Loadable Software} - Refers to software that can be loaded without removing the system or equipment from its installation.
\end{itemize}

In essence, the \emph{DO-178C} document goes beyond just failure conditions, encompassing these diverse aspects of software considerations within the context of the system lifecycle process.

\subsection{DO-178C and related Supplements}
\label{do178candsupplements}
\begin{figure}[htbp]
\centerline{\includegraphics[width=0.6\textwidth]{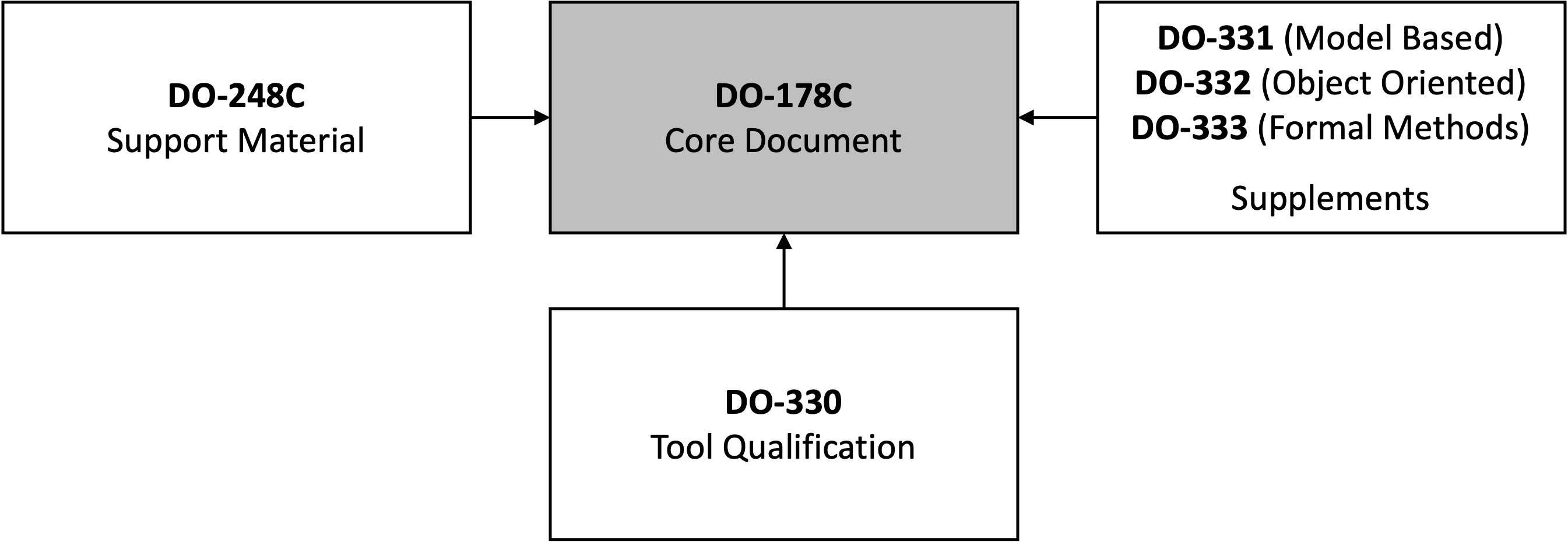}}
\caption{\emph{DO-178C} and related documents, adapted from \cite{Rierson2013}.}
\label{fig3}
\end{figure}

The \emph{DO-178C} standard guides the aerospace community in developing with an acceptable level of confidence the software parts of airborne systems and equipment that must comply with the requirements of the standards and documents. It describes the complete systems lifecycle process, including the safety assessment and validation processes; however, it does not intend to describe the certification process for which we must refer to applicable regulations and guidance material issued by the certification authorities \cite{DO-178C}. Figure \ref{fig3} shows the relationship between the \emph{DO-178C} and the related supplements. As we can observe, the \emph{DO-178C} is what we can call the core document.

Each one of the \emph{DO-178C} supplements provides additional information on key topics, namely \cite{DO-178C, Rierson2013}:
\begin{itemize}
    \item \textbf{\emph{DO-248C} (Supporting Information for \emph{DO-178C} and \emph{DO-278A})} -- Collection of Frequently Asked Questions (FAQs) and Discussion Papers (DPs) with applications of \emph{DO-178C} and \emph{DO-278A} in the safety assurance of software for aircraft \cite{DO-248C}. 
    \item \textbf{\emph{DO-330} (Software Tool Qualification Considerations)} -- Standalone document referenced by the \emph{DO-178C} that provides guidance on tools qualification \cite{DO-330}.
    \item \textbf{\emph{DO-331} (Model-Based Development and Verification Supplement to \emph{DO-178C} and \emph{DO-278A})} -- Provides guidance for model-based development and verification used as part of the software lifecycle. Furthermore, it includes the outputs that would be using models and the verification evidence that could be derived from them. Therefore, this supplement also applies to the models developed in the system process that define software requirements or software architecture \cite{DO-331}.
    \item \textbf{\emph{DO-332} (Object-Oriented Technology and Related Techniques to \emph{DO-178C} and \emph{DO-278A})} -- Provides guidance when object-oriented technology or related techniques are used as part of the software development lifecycle (development and verification) \cite{DO-332}.
    \item \textbf{\emph{DO-333} (Formal Methods Supplement to \emph{DO-178C} and \emph{DO-278A})} -- Provides guidance when formal methods are used as part of a software lifecycle \cite{DO-333}.
\end{itemize}

The current focus of this study is the \emph{DO-178C}; however, these other documents might be considered in the future. \emph{DO-178C} itself comprises twelve sections, two annexes and two appendixes briefly described in Subsection\ref{do178cstructure} \cite{DO-178C, Rierson2013}.

The main purpose of the \emph{DO-178C} standard is not to explain "how to develop" but rather to "support the development" of software for airborne systems and equipment to ensure a level of confidence in safety that complies with specific airworthiness requirements/objectives \cite{DO-178C}. This demonstrates that there are no inherent constraints or restrictions within the document that forbid the adoption of Agile methods, contrary to the prevailing industry opinion that the traditional waterfall method is mandatory.

\subsection{DO-178C Document Structure}
\label{do178cstructure}
Having a better understanding of how the \emph{DO-178C} relates to the other documents and supplements (Section \ref{aerospacestandardsanddocuments} and Subsection \ref{do178candsupplements}), we believe it is the right moment to advance and understand the \emph{DO-178C} structure.

The document is composed of 12 sections, briefly described below \cite{DO-178C, Rierson2013}:
\begin{itemize}
    \item \textbf{Section 1} -- document overview. Explains the purpose and scope of the document, its relationship to other documents and how to use it.
    \item \textbf{Section 2} -- provides details about the system framework related to software development (such as: information flow between systems and software lifecycle process, system and software assessment process, among others); it is aligned with \emph{ARP4745A}.
    \item \textbf{Section 3} -- software lifecycle process overview of planning, development (requirements, design, code and integration), verification, software configuration management, software quality assurance and certification liaison.
    \item \textbf{Section 4} -- explains the planning process, objectives and activities. One example of a planning process is the software lifecycle environment from development to testing.
    \item \textbf{Section 5} -- provides the necessary details for the software development process objectives and activities that include requirements, design, coding, integration and traceability phases.
    \item \textbf{Section 6} -- verification is not just testing but a technical assessment of the outputs. Verification is vital in software development assurance since over half of the \emph{DO-178C} objectives are verification objectives. The process begins with the verification of the plans, and it goes until reporting and review of verification results.
    \item \textbf{Section 7} -- explains the change configuration management process. The process includes objectives and activities like traceability reporting, control and review.
    \item \textbf{Section 8} -- provides guidance for the software quality assurance process, what are the objectives and activities.
    \item \textbf{Section 9 and 10} -- summarize the certification liaison and the certification process.
    \item \textbf{Section 11} -- identifies the software lifecycle outputs (data) generated while planning and verifying the software.
    \item \textbf{Section 12} -- includes guidance for some additional considerations, including modifications to previously developed software, tool qualification, alternative methods, software reliability models and product service history.
\end{itemize}

The \emph{DO-178C} document also contains annexes A and B, and appendixes A and B covering \cite{DO-178C, Rierson2013}:
\begin{itemize}
    \item \textbf{Annex A} -- contains tables for each process with the 71 formal objectives and outputs by software level (e.g., independence requirements for the objective, output to be generated for safety, the objective and the amount of configuration control required on the output).
    \item \textbf{Annex B} -- includes the acronyms and glossary of terms used in the \emph{DO-178C} standard.
    \item \textbf{Appendix A} -- provides the background of the \emph{DO-178} documents and highlights the differences between the previous (\emph{DO-178B}) and the current (\emph{DO-178C}) standards.
    \item \textbf{Appendix B} -- contains committee membership.
\end{itemize}

Once more, the purpose of the \emph{DO-178C} standard is not to explain the "how", instead and quoting \cite{DO-178C} is to:

\begin{quotation}
\noindent The purpose of this document is to provide guidance for the production of software for airborne systems and equipment that performs its intended function with a level of confidence in safety that complies with airworthiness requirements.
\end{quotation}

This conclusion emphasizes the absence of intrinsic limitations or restrictions within the document that would deny the adoption of Agile methods, challenging the general industry opinion that the traditional waterfall method is mandatory.

\subsection{DO-178C Five Plans}
\label{DO-178CFivePlans}
Subsection \ref{do178cstructure} provided valuable insights into the \emph{DO-178C} structure, clarifying the significance of each of its twelve sections. Turning our focus to section eleven within the \emph{DO-178C} document, it strongly recommends creating five plans with specific information for any DAL level, which is referred to as the "Software level" in the \emph{DO-178C} (last version published in 2011). These plans can be summarized as follows \cite{DO-178C, Rierson2013}:
\begin{itemize}
    \item \textbf{Plan for Software Aspects of Certification (PSAC)} -- This is the preliminary plan submitted to the certification authority. It is used to document the agreements with the certification authority and to determine if the proposed software lifecycle corresponds with the rigour required for the software's DAL being developed by the applicant.
    \item \textbf{Software Development Plan (SDP)} -- It describes the software development procedures and lifecycle to satisfy the process objectives to guide the development team and ensure compliance. Furthermore, it describes the development environment and the standards to be used (requirements, design and coding). The SDP provides the necessary details to ensure proper implementation of the software lifecycle process, while the PSAC only contains a summary.
    \item \textbf{Software Verification Plan (SVP)} -- It describes the software development procedures to be used and guides the verifiers through the verification process, ensuring compliance with the \emph{DO-178C} objectives. Some examples are organisational responsibilities, independence, verification methods and environment.
    \item \textbf{Software Quality Assurance Plan (SQAP)} -- Establishes the plan and methods to achieve the \emph{DO-178C} document objectives, plans, and software quality assurance standards.
    \item \textbf{Software Configuration Management Plan (SCMP)} -- Establishes the software configuration management procedures, activities, and environment to be used throughout the software development and verification. The plan includes configuration identification, baselines and traceability, problem reporting, change control, and change review.
\end{itemize}

Each of the five plans (PSAC, SDP, SVP, SQAP, and SCMP) outlines the processes that must be implemented and applied to the project's outputs \cite{DO-178C, Rierson2013}. These processes are meticulously defined within these plans right at the inception of any aerospace project, even before the specification of requirements, encompassing both system and software aspects within the aerospace system and software development context. It is crucial to emphasize that the \emph{DO-178C} document deliberately refrains from imposing or constraining the selection of methods to be employed and detailed in these plans. This flexibility empowers project stakeholders to opt for the method that best aligns with their requirements, including Agile methods. The pivotal factor to consider when choosing a method lies in ensuring that it instils a level of safety confidence that meets the specific airworthiness requirements and objectives tailored to the particular DAL, as elaborated in Section \ref{failureconditioncategorylevel}.

In summary, the \emph{DO-178C} outlines a structured and rigorous approach to software development for safety-critical aerospace systems. While the text does not explicitly state that \emph{DO-178C} restricts the usage of Agile software development, the \emph{DO-178C} places significant emphasis on documentation, as outlined in the five plans considering the various aspects of the software development and verification process. \emph{DO-178C} mandates rigorous verification and quality assurance processes, which include maintaining independence between development and verification activities. While Agile practices often promote continuous integration and delivery, alternative approaches are available to ensure role independence while still adhering to verification and quality assurance requirements. These approaches may include practices like Test-Driven Development (TDD) and Pair Programming (PP)\footnote{Test-Driven Development (TDD) and Pair Programming (PP) are two of the twelve practices that make up the Agile method known as Extreme Programming (XP), introduced by Kent Beck \cite{Beck1999, Beck2004}.} \cite{VanderLeest2009}. \emph{DO-178C} requires adherence to specific predefined standards for requirements, design, and coding. While Agile teams typically enjoy flexibility in selecting their own, including the tools to be used, it is essential to note that when defining the standards within the plans, an agreement can be achieved with the teams to adhere to the specified guidelines. Lastly, \emph{DO-178C} requires tailoring its processes according to the DAL of the software being developed, indicating different levels of rigour for different DALs. In this context, an Agile method that aligns with the specific DAL may be selected. However, it is essential to acknowledge that customizations might be necessary to address the particular DAL requirements while ensuring compliance with \emph{DO-178C}. Considering these factors, we conclude that \emph{DO-178C}'s emphasis on detailed planning, documentation, and adherence to specific standards makes it challenging to adopt Agile and implement a particular Agile method without significant modifications. While it is not impossible to use Agile in a \emph{DO-178C}-compliant project, it would likely require tailoring the Agile method selected and respective practices to meet the rigorous safety and certification requirements of \emph{DO-178C}. Ultimately, the choice of development, including an Agile method, should be made carefully, considering the specific needs and constraints of the project and the regulatory requirements of \emph{DO-178C}.

\subsection{Software Development Standards and Plans Outputs}
\label{SoftwareDevelopmentStandardsandPlansOutputs}
Along with the five plans described in the previous subsection (Subsection \ref{DO-178CFivePlans}), \emph{DO-178C} also requires adherence to three additional standards \cite{DO-178C, Rierson2013}: Software Requirement Standard (SRS), Software Design Standard (SDS), and Software Coding Standard (SCS). While these standards can draw inspiration from industry-wide standards, they must be defined individually and customized for project-specific needs. These documents establish a set of rules and constraints for the project team, facilitating control over functionality and safety features.

While it is essential to recognize that \emph{DO-178C} mandates compliance with specific predefined standards for requirements, design, and coding, and Agile teams generally have the flexibility to select their own standards and tools, it is equally crucial to emphasize that no inherent limitation exists within Agile methods that prevents alignment with the specified guidelines when defining standards within the plans. Moreover, there are no inherent constraints that prevent Agile teams from adhering to these guidelines, as agreements can be reached with the teams to ensure compliance with those requirements.

\begin{figure}[htbp]
\centerline{\includegraphics[width=1\textwidth]{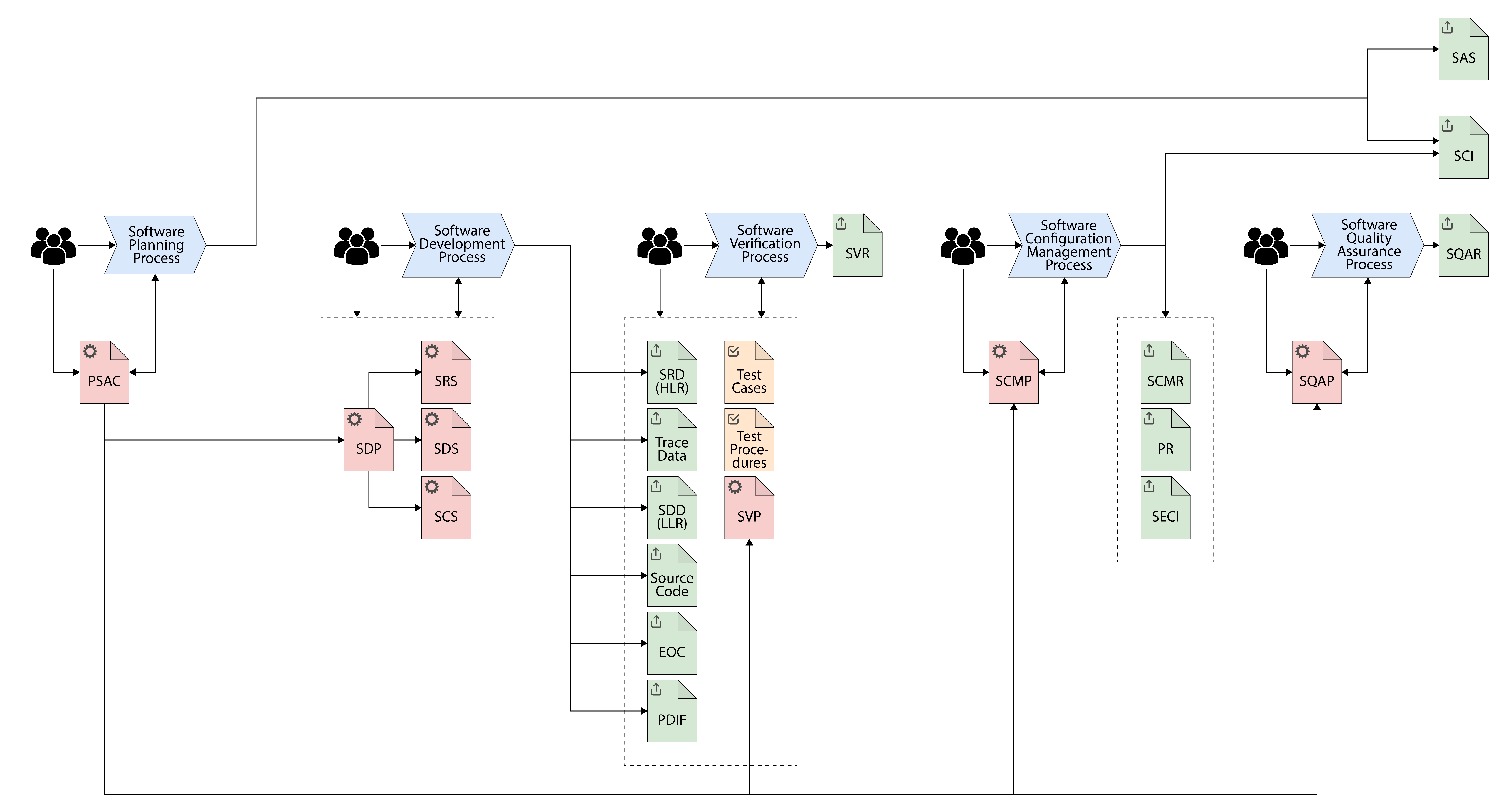}}
\caption{Plans and processes correlation with their corresponding outputs.}
\label{fig4}
\end{figure}

Returning our attention to the outputs of the five plans, there is a strong correlation between the plans, processes, standards, and outputs, all collectively contributing to achieving compliance with \emph{DO-178C} objectives \cite{DO-178C, Rierson2013}, as illustrated in Figure \ref{fig4} and described below.

\vspace{0.5cm}

\begin{itemize}
    \item \textbf{Software Accomplishment Summary (SAS)} -- summarises the actual versus planned (as specified in PSAC) activities and results for the project. The certification authority uses it to overview the software aspects of certification.
    \item \textbf{Software Configuration Index (SCI)} -- identifies the software configuration and can contain a set of configuration items.
    \item \textbf{Software Requirements Data (SRD)} -- defines the high-level requirements (HLR), including the derived requirements.
    \item \textbf{Trace Data} -- establishes the associations between lifecycle data items and contents. Trace Data demonstrates bi-directional associations between:
    \begin{itemize}
        \item System requirements allocated to software and HLR;
        \item HLR and low-level requirements (LLR);
        \item LLR and source code;
        \item Software requirements and test cases;
        \item Test cases and test procedures;
        \item Test procedures and test results.
    \end{itemize}
    \item \textbf{Software Design Description (SDD)} -- defines the software architecture and the LLR that will satisfy the HLR.
    \item \textbf{Source Code} -- consists of written code in the source language.
    \item \textbf{Executable Object Code (EOC)} -- consists of code usable by the target computer processing unit and is, therefore, software loadable into the hardware or system.
    \item \textbf{Parameter Data Item File (PDIF)} -- consists of usable data by the target computer processing unit.
    \item \textbf{Software Verification Results (SVR)} -- results produced by the software verification activities. It is used as evidence to support the system processes assessment.
    \item \textbf{Software Configuration Management Records (SCMR)} -- results of the software configuration management process’s activities. Examples are configuration identification lists, baselines, change history reports, and release records.
    \item \textbf{Problem Reports (PR)} -- are a means to identify and record software product anomalous behaviour resolution, process non-compliances with software plans and standards, and deficiencies in software lifecycle data.
    \item \textbf{Software lifecycle Environment Configuration Index (SECI)} -- identifies the software lifecycle environment’s configuration. This index is written to reproduce the hardware and software lifecycle environment for software regeneration, re-verification, or modification.
    \item \textbf{Software Quality Assurance Records (SQAR)} -- SQA records results from audit reports, meeting minutes, authorized process deviation, software conformity review records, and others.
\end{itemize}

In summary, as illustrated in Figure \ref{fig4}, the production of these outputs occurs at different stages throughout the software development lifecycle, as outlined in the project plans. These outputs ensure adherence to \emph{DO-178C} objectives. The timing of output generation can vary depending on the selected software development method, the project's lifecycle, and the development phase in progress. The exact timing also depends on project-specific considerations, the chosen development method, and the DAL assigned to the software under development. Achieving \emph{DO-178C} compliance requires a rigorous approach encompassing comprehensive planning, meticulous documentation, and verification activities at each phase, guaranteeing that safety-critical aerospace software aligns with its intended objectives. Considering that the \emph{DO-178C} does not restrict the use of specific development methods, including Agile methodologies.

Furthermore, Agile, as per the Agile Manifesto, is characterized by an iterative and incremental approach focusing on continuous delivery and client satisfaction \cite{Beck2001}. Transitioning to Agile in this context is feasible without harsh changes to existing processes \cite{VanderLeest2009, SilvaCardoso2022}. Instead, it can be accomplished by defining a software development process in the plans (Subsection \ref{DO-178CFivePlans}) that is tailored and incorporates some Agile practices, aligning them with the specific DAL requirements to ensure compliance with \emph{DO-178C}. This results in a more streamlined process and the potential to reuse outputs generated during different lifecycle phases, including documentation, iteratively and incrementally.

\subsection{Software Development Process}
\label{SoftwareDevelopmentProcess}
As mentioned earlier, both the PSAC and the SDP must provide a comprehensive description of the software development process to be used. It is important to remark that \emph{DO-178C} does not prescribe specific methods, neither does it restrict Agile methods. Nevertheless, following DO-178C guidelines, the software development process consists of four subprocesses \cite{DO-178C, Rierson2013}: the Software Requirements Process, Software Design Process, Software Coding Process, and Integration Process. In this publication, we do not aim to define or describe the detailed methods, including an Agile method tailoring for each process. They are individually defined and customized to address project-specific requirements within the PSAC and SDP plans, which are established before starting any development. Our future work may explore the incorporation of Agile methods with the implementation of the required tailoring into these processes and always maintaining compliance with the \emph{DO-178C} objectives.

\begin{figure}[htbp]
\centerline{\includegraphics[width=0.65\textwidth]{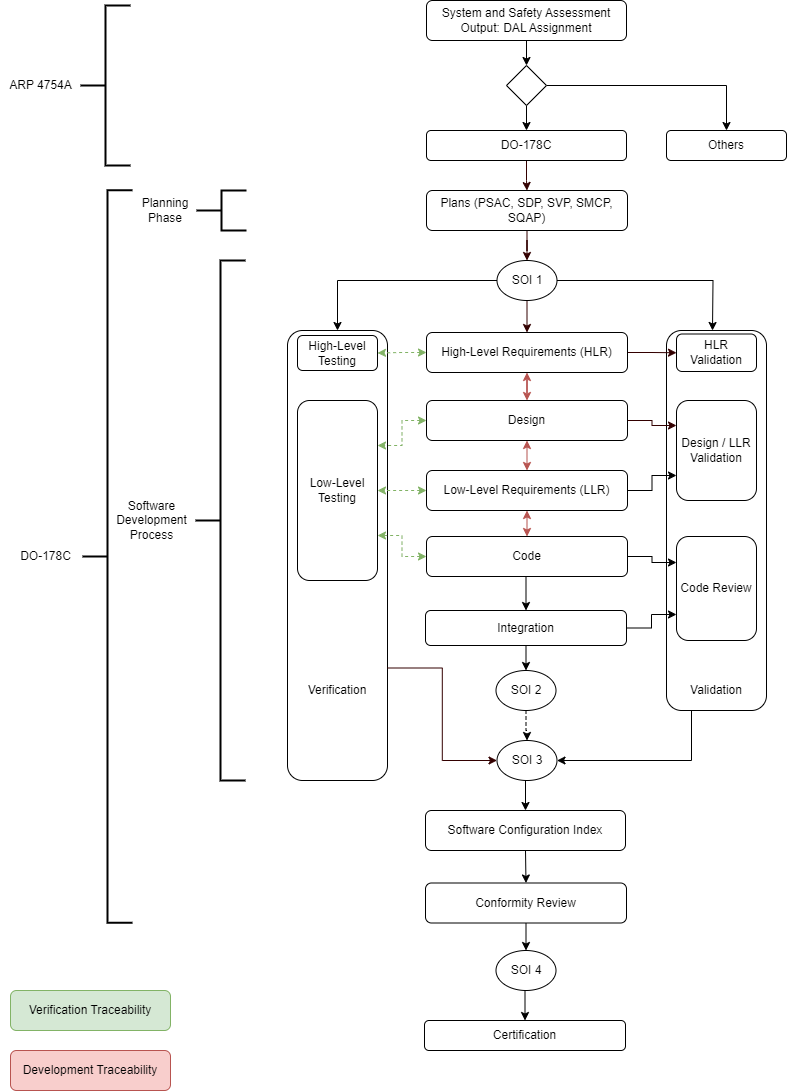}}
\caption{\emph{DO-178C} Software Development Process, adapted from \cite{DO-178C}.}
\label{fig5}
\end{figure}

Nonetheless, it is worth noting that certain aerospace companies have achieved a high level of maturity by relying on predefined plans and software development processes (traditional waterfall method), necessitating only minor tailoring to adapt to the specific needs of new projects. To provide a comprehensive understanding of the relationship between plans, processes, and V\&V activities and how they interconnect, we have drawn upon the insights gained from our analysis of \emph{DO-178C}. Figure \ref{fig5} illustrates the outcome of this exercise. Our primary objective has been to identify and describe the essential steps from the planning phase to what we consider as certifiable or certification-ready. Furthermore, to ensure successful certification, it is crucial to grasp the significance of the State of Involvement (SOI) within the process. The certification authority outlines SOI assessments/audit events at various project stages \cite{Rierson2013}.

The \emph{DO-178C} standard does not define or indicate the certification authority involvement level in a certification process. However, the author in \cite{Rierson2013} states that the International Certification Authorities Software Team (CAST) identified a software compliance assessment approach indicating four intervention points throughout the project: planning, development, test and final \cite{DO-178C, Rierson2013}.

As stated by \cite{Rierson2013}:
\begin{itemize}
    \item \textbf{SOI 1} -- audit that happens after the plans and standards have been reviewed and baselined.
    \item \textbf{SOI 2} -- audit that happens after at least half of the code has been developed and reviewed.
    \item \textbf{SOI 3} -- audit usually happens after at least half of the test cases and procedures are developed and reviewed.
    \item \textbf{SOI 4} -- audit happens after any issues from the other SOI audits are closed, and the verification results, SCI and SAS are reviewed and baselined.
\end{itemize}

Figure \ref{fig5} shows how the software requirements process uses the system lifecycle processes’ outputs to develop HLR; the software design process uses the HLR to redefine and develop the software architecture and the LLR that can be used to implement the source code. The mapping also shows how integration and V\&V relate with HLR, Design, LLR and source code. Furthermore, Figure \ref{fig5} shows when each SOI event should achieve successful V\&V and certification-ready status. However, the \emph{DO-178C} does not specify any software development method such as Waterfall, or any other method to be used by any process; it only requires that what is described in the plans is followed and the production of the needed evidence to achieve certification \cite{DO-178C, Rierson2013, VanceHildermanandTonyBaghai2007}.

\subsection{Traceability, Independence and Change Control}
\label{TraceabilityIndependenceandChangeControl}
The \emph{DO-178C} standard identifies traceability, independence and change control activities as highly important to achieving successful certification.

Following the \emph{DO-178C} standard, traceability, independence and change control can be summarised as \cite{DO-178C}:
\begin{itemize}
    \item \textbf{Traceability} -- is the mapping between process outputs, an output and its originating process, or between a requirement and its implementation.
    \item \textbf{Independence} -- separation of responsibilities to guarantee an objective evaluation. They are usually achieved by having the verification activity performed by person(s) or tools independent from the developer of the item being verified.
    \item \textbf{Change control} -- managing and controlling changes to configuration items and baselines after their formal establishment.
\end{itemize}

In summary, an Agile method can be customized by integrating specific Agile practices when defining the software development process in the PSAC and SDP. This method should be defined and implemented in alignment with the requirements and objectives of \emph{DO-178C} to ensure compliance, specifically when addressing traceability, independence, and change control requirements. For traceability, Product Backlog Items (PBIs)\footnote{ PBI is a smaller requirement increment of work that can be completed during a specific established timeframe \cite{Schwaber2020}.} can establish clear links between user requirements and development tasks or features. Each PBI can comprehensively outline the requirement and its corresponding implementation, ensuring a transparent connection between requirements and their successful implementation. Acceptance criteria, specifying the conditions for requirement fulfilment, further enhance traceability. To achieve independence, Agile practices like cross-functional teams promote collaboration among members with diverse roles, ensuring independence. This structure inherently fosters objective evaluations, particularly during verification, where team members not involved in initial development can provide impartial assessments. Code reviews within the Agile method also ensure independent evaluations of code quality and alignment with requirements.

Regarding change control, Agile teams effectively manage changes through continuous refinement and prioritization of PBIs. This approach allows teams to accommodate new requirements while maintaining traceability. In addition, Agile teams usually employ version control systems to track and manage changes to configuration items systematically, ensuring comprehensive documentation and management. By integrating these practices, the process can successfully address traceability, independence, and change control requirements while upholding the requirements and objectives outlined in \emph{DO-178C}.

\subsection{DO-178C Objectives and Activities}
\label{do178cactivities}
In Subsection \ref{failureconditioncategorylevel}, Table \ref{tab4} provides a summary of the distinct DAL levels, going from the highest level of criticality (level A) to the lowest (level E). The table also presents the number of objectives and verification objectives that must be adhered to according to the \emph{DO-178C} document for each level.
Moving forward, Subsection \ref{DO-178CFivePlans} outlines the essential five plans mandated by the \emph{DO-178C} document. This subsection delves into the intricate relationships between these plans, elucidates their individual processes, and delineates the resultant outputs they yield.
Moreover, Subsection \ref{SoftwareDevelopmentProcess} introduces Figure \ref{fig5}, which visually elucidates the fundamental stages spanning from the initial planning phase to the attainment of a state that qualifies as either certifiable or ready for certification.

As previously discussed in Subsection \ref{do178cstructure}, Annex A of the \emph{DO-178C} document contains tables that elaborate on the objectives, verification objectives activities, and outputs for various software levels \cite{DO-178C}. These tables pertain to the following processes:
\begin{itemize}
    \item Software Planning Process (Table-1, Page 96, \cite{DO-178C});
    \item Software Development Processes (Table-2, Page 97, \cite{DO-178C});
    \item Verification of Outputs of Software Requirements Process (Table-3, Page 98, \cite{DO-178C});
    \item Verification of Outputs of Software Design Process (Table-4, Page 99, \cite{DO-178C});
    \item Verification of Outputs of Software Coding \& Integration Processes (Table-5, Page~100, \cite{DO-178C});
    \item Testing of Outputs of Integration Process (Table-6, Page 101, \cite{DO-178C});
    \item Verification of Verification Process Results (Table-7, Page 102, \cite{DO-178C});
    \item Software Configuration Management Process (Table-8, Page 103, \cite{DO-178C});
    \item Software Quality Assurance Process (Table-9, Page 104, \cite{DO-178C});
    \item Certification Liaison Process (Table-10, Page 105, \cite{DO-178C}).
\end{itemize} 

Furthermore, It is important to note that the \emph{DO-178C} document emphasizes that these tables, elucidating objectives, activities, and outputs, should not be regarded as checklists, as they do not encompass the entirety of compliance aspects stipulated by the standard \cite{DO-178C}. Due to the above reason, we intend not to detail each activity mandated by the \emph{DO-178C} document exhaustively, but rather to indicate the locations where these details can be sourced in the future for each respective process.

Moreover, listing all objectives and verification objectives by each DAL is impractical due to the extensive and complex nature of the information. For the complete list of objectives and verification objectives, it is recommended to refer to the \emph{DO-178C} document itself, which provides comprehensive guidance on the requirements for each level. The document is extensive and complex, designed to ensure the highest levels of safety and reliability in airborne software systems.

\section{Conclusions}
\label{conclusions}
In conclusion, the \emph{DO-178C} document and its supplements represent a commitment to aerospace software safety. As aerospace systems rely heavily on software to enhance efficiency and functionality, the guidelines outlined in the \emph{DO-178C} document have emerged as a critical influence in shaping aerospace domain software development. At its core, the \emph{DO-178C} document is not intended to be a rigid and prescriptive rulebook for software development. Instead, its primary purpose is to provide comprehensive guidance for software production within the aerospace domain. This approach acknowledges the dynamic nature of software development nowadays while setting a clear path toward ensuring safety and reliability. By avoiding a one-size-fits-all method, the \emph{DO-178C} document encourages adaptability and innovation in the pursuit of robust software solutions. In other words, it does not impose restrictions on the adoption of Agile methods, a finding that we present in this paper that challenges the prevailing industry notion that the traditional waterfall method is mandatory. Central to the \emph{DO-178C} are the five plans: the PSAC, SDP, SVP, SQAP, and SCMP. These plans collectively lay the groundwork for the software development process, delineating a course to prioritise traceability, independence, and change control. This holistic approach ensures that software development is not merely a technical endeavour but a rigorously documented and audited journey that leaves no room for ambiguity. Under the \emph{DO-178C}, the software development process encompasses four integral subprocesses: Software Requirements Process, Software Design Process, Software Coding Process, and Integration Process.

Figure \ref{fig5} shows the relationship between these subprocesses, plans, and safety certification objectives. The SOI audits, dispersed throughout the development lifecycle, underscore the incremental nature of verification and the critical role of continuous assessment in achieving the highest airworthiness standards. The \emph{DO-178C} document being a reference for industry collaboration, it embodies the collective dedication to safety within the aerospace domain. Offering guidance rather than a strict software development method encourages an environment where developers, engineers, and stakeholders can tailor their software development practices to address unique challenges and objectives, leaving space for future adaptations resulting from an ever-evolving industry landscape. While  \emph{DO-178C} does not explicitly restrict Agile methods, its rigorous framework places significant emphasis on documentation, independence between development and verification activities, and adherence to predefined standards, aspects that are not generally taken into account in Agile methods. By incorporating in Agile development practices such as traceability through Product PBIs, cross-functional teams for independence, and change control, it is possible to use Agile methods within the \emph{DO-178C} compliance context. Agile can offer benefits like continuous integration and delivery while maintaining compliance, but it must align with \emph{DO-178C}'s requirements and objectives. Achieving a successful integration of Agile practices into \emph{DO-178C} involves tailoring the selected Agile method and practices, ensuring that they address the specific needs and safety certification requirements of the project. Ultimately, the choice of development approach, including Agile, should be made thoughtfully, considering project specifics and the regulatory requirements of \emph{DO-178C}.

Given the adaptable nature of the \emph{DO-178C} document regarding software development methods, future research could focus on two main areas:
\begin{itemize}
    \item Evaluating Established Methods: A comprehensive analysis of overall methods, including a critical examination of the Waterfall method, considering scientific research and industry experiences.
    \item Exploring New Approaches: Investigating emerging software development methods from the Agile field that can accommodate the dynamic demands of the aerospace domain. The goal is to identify methods that optimize efficiency, safety, and compliance within the \emph{DO-178C} document.
\end{itemize}

Both research lines could contribute to increasing the domain knowledge and refining aerospace software practices, fostering a balance between innovation and safety in this rapidly evolving industry.


\bibliographystyle{unsrtnat}


\begin{thebibliography}{1}




\bibitem{Bowen1993}Bowen, J. \& Stavridou, V. Safety-critical systems, formal methods and standards. {\em Software Engineering Journal}. \textbf{8}, 189 (1993)

\bibitem{Cullyer1993}Cullyer, J. Safety critical systems. {\em Microprocessors And Microsystems}. \textbf{17}, 2 (1993)

\bibitem{Knight2002}Knight, J. Safety critical systems: Challenges and directions. {\em Proceedings - International Conference On Software Engineering}. pp. 547-550 (2002)

\bibitem{Myklebust2018}Myklebust, T. \& St{\aa}lhane, T. The agile safety case. {\em The Agile Safety Case}. pp. 1-235 (2018,1)

\bibitem{Mathur2010}Mathur, S. \& Malik, S. Advancements in the V-Model. {\em ©2010 International Journal Of Computer Applications}. \textbf{1} pp. 975-8887 (2010)

\bibitem{Russo2016}Russo, S. \& Scippacercola, F. Model-Based Software Engineering and Certification: Some Open Issues. {\em Proceedings - 2016 IEEE 27th International Symposium On Software Reliability Engineering Workshops, ISSREW 2016}. pp. 237-240 (2016,12)

\bibitem{DO-178C}DO-178C, Software Considerations in Airborne Systems and Equipment Certification. {\em RTCA}. (2011)

\bibitem{Rierson2013}Rierson, L. Developing Safety-Critical Software: A Practical Guide for Aviation Software and DO-178C Compliance. (CRC Press,2013)

\bibitem{VanceHildermanandTonyBaghai2007}Vance Hilderman and Tony Baghai Avionics Certification: A Complete Guide to DO-178 (Software), DO-254 (Hardware). (2007)

\bibitem{DO-254}DO-254, Design Assurance Guidance for Airborne Electronic Hardware. {\em RTCA}. (2000)

\bibitem{ARP4754A}ARP4754A, Guidelines for Development of Civil Aircraft and Systems. {\em SAE International, December}. (2010)

\bibitem{ARP4761}ARP4761, Guidelines and methods for conducting the safety assessment process on civil airborne systems and equipment. {\em SAE International, December}. (1996)

\bibitem{DO-297}DO-297, Integrated Modular Avionics (IMA) Development Guidance and Certification Considerations. {\em RTCA}. (2005)

\bibitem{ARP5150A}ARP5150A, Safety Assessment of Transport Airplanes in Commercial Service. {\em SAE International}. (2013)

\bibitem{ARP5151A}ARP5151A, Safety Assessment of General Aviation Airplanes and Rotorcraft in Commercial Service. {\em SAE International}. (2013)

\bibitem{Singh2011}Singh, A. RTCA DO-178B (EUROCAE ED-12B). (2011), https://www.researchgate.net/publication/315023556

\bibitem{McHale2001}McHale, J. Real-time operating system vendors rush to comply with DO-178B | Military \& Aerospace Electronics. (2001), https://www.militaryaerospace.com/computers/article/16710702/realtime-operating-system-vendors-rush-to-comply-with-do178b

\bibitem{DO-248C}DO-248C, Supporting Information for DO-178C and DO-278A. {\em RTCA}. (2011)

\bibitem{DO-330}DO-330, Software Tool Qualification Considerations. {\em RTCA}. (2011)

\bibitem{DO-331}DO-331, Model-Based Development and Verification Supplement to DO-178C and DO-278A. {\em RTCA}. (2011)

\bibitem{DO-332}DO-332, Object-Oriented Technology and Related Techniques to DO-178C and DO-278A. {\em RTCA}. (2011)

\bibitem{DO-333}DO-333, Formal Methods Supplement to DO-178C and DO-278A. {\em RTCA}. (2011)

\bibitem{VanderLeest2009}VanderLeest, S. \& Buter, A. Escape the waterfall: agile for aerospace. {\em 2009 IEEE/AIAA 28th Digital Avionics Systems Conference. DASC 2009}. (2009)

\bibitem{Beck1999}Beck, K. Extreme Programming Explained: Embrace Change. (Addison-Wesley Longman Publishing Co., Inc.,1999)

\bibitem{Beck2004}Beck, K. \& Andres, C. Extreme Programming Explained: Embrace Change (2nd Edition). (Addison-Wesley Professional,2004)

\bibitem{Beck2001}Beck, K., Beedle, M., Van Bennekum, A., Cockburn, A., Cunningham, W., Fowler, M., Grenning, J., Highsmith, J., Hunt, A., Jeffries, R., Kern, J., Marick, B., C. Martin, R., Mellor, S., Schwaber, K., Sutherland, J. \& Thomas, D. Manifesto for Agile Software Development, (URL http://agilemanifesto.org/).  (2001)

\bibitem{SilvaCardoso2022}Silva Cardoso Rodrigues, J., Ferreira Ribeiro, J. \& Aguiar, A. Improving Documentation Agility in Safety-Critical Software Systems Development For Aerospace. {\em 2022 IEEE International Symposium On Software Reliability Engineering Workshops (ISSREW)}. pp. 222-229 (2022)

\bibitem{Schwaber2020}Schwaber, K. \& Sutherland, J. Scrum Guide V7.  (2020)

\end{thebibliography}

\end{document}